\documentclass[preprint,12pt]{elsarticle}
\usepackage[pdftex]{hyperref}
\usepackage{graphicx}

\usepackage{amssymb}
\usepackage[fleqn]{amsmath}
\usepackage{bm}
\usepackage{amsbsy}
\usepackage{lineno}
\usepackage{physics}
\usepackage{mathtools}
\usepackage{xparse}
\usepackage{accents}

\usepackage{natbib}

\begin{document}

\title{Revisiting the Hamilton theory for second order Lagrangian}


\author{Israel A. Gonz\'alez Medina}
%

\address{Instituto Superior de Tecnologias y Ciencias Aplicadas.\\ Universidad de la Habana. \\Cuba}
\ead{israelariel.gonzalezmedina@gmail.com}


\date{\today}

\begin{abstract}
The Hamilton theories for higher orders classical Lagrange functions result on a well known Ostrogradski’s instabilities. In this work, we propose a different definition for the second order canonical momentum and obtain a new set of second order's Hamilton equations. The identity transformation introduces a new set of constraints depending only on the set of velocities of all particles and removing the Ostrogradsky’s instability. The evolution of the system identifies a new set of canonical variables as the poles of the constraints. The second order momentum shows to be the generator for the negative displacement of poles of such constraints.  The momentum first order momentum remains as the generator for the displacement of the coordinate.
\end{abstract}


\maketitle

\section*{Introduction}
Theories with higher order Lagrangians have been explored along the evolution of physics. They are proposed as solutions from alternative theories of gravitation\cite{Langlois:2015cwa} to fundamental particle theories \cite{Woodard:2015zca}. Individually, theories of second-order Lagrangian rise a remarkable interest because they are renormalizable \cite{PhysRevD.16.953} in four dimensions. 

However, while the Lagrangian has no physical meaning, while Hamiltonian provide more in-depth knowledge of the classical mechanic structure and sets equals status for coordinates and momenta as independent variables. Also, Hamiltonian is related to essential system features such as energy and also provide significant relations between symmetry and conservation laws. Because of that a problem arise when second order Lagrangians problems are described with under the Hamilton approach. Indeed, as a consequence of a theorem of Mikhail Ostrogradsky, nondegenerate Lagrangians with higher order time derivatives lead to ghost-like instabilities, also known as Ostrogradski instabilities \cite{Ostrogradsky:1850fid}. This behavior results in a linearly unstable Hamiltonians in such a way that they cannot be eliminated by partial integration. By far, this the the greatest restriction is the obstacle to including higher time derivatives in the canonical  formalism of nondegenerate higher derivative Lagrangians.

In our case, the main motivation is to find the Hamilton equations for the second order Lagrangians obtained after considering the mass of the particles as variables magnitudes. The Lagrangians form part of a more general proposal for the quantum description of the isolated particle systems with $n$-particles with variable masses and connected by a field with variable form ($n$-VMVF systems). The proposal revise the Lagrange theory for the referred physical systems whose start point is the noncompliance of the second Newton law for variable mass. This fact lead to the proposal for an extension of the D'Alembert principle. One consequence of this analysis is that particle can be isolated when mass is a variable quantity.

The pre-print version of the entire proposal for the construction of the Quantum Theory including masses and field as unknown functions can be found in the article entitled ``\href{http://arxiv.org/abs/1811.12175}{A new proposal for a quantum theory for isolated n-particle systems with variable masses connected by a field with variable form}'' \cite{Israel:1811.12175}. The pre-print version of the revision of the Lagrange theory for those systems can be found in the article entitled ``\href{http://arxiv.org/abs/1903.04916}{Revisiting the Lagrange theory for isolated n-particle systems with variable masses connected by an unknown field}'' \cite{Israel:1903.04916}.

This work covers the second item of the methodology of developing the Hamilton theory for second order Lagrangian of those systems and it propose a different construction for it removing the referred instability.

\section{The Ostrogradsky's construction of the Hamiltonian for the second order Lagrangian}
For the treatment of higher derivative systems, Ostrogradsky generalizes the construction of the Hamilton function \cite{Ostrogradsky:1850fid}. In this section, we briefly expose his main ideas for the second order Lagrangian. 

Let us consider the extended Lagrange equation for a single particle \cite{Courant53physics}:
\begin{equation}
\frac{d^2}{d t^2}\Big( \frac{\partial L}{\partial \ddot{q}}\Big) - \frac{d}{dt}\Big( \frac{\partial L}{\partial \dot{q}}\Big) + \frac{\partial L}{\partial q} =0
\end{equation}
describing a system whose Lagrange function $L(q,\dot{q},\ddot{q})$ non degenerately depends on $\ddot{q}$, which implies that the Hessian $\frac{\partial^2 L}{\partial \ddot{q}^2}\neq 0$. In this case, the four derivative's term can be expressed as a function of the others as
\begin{equation}
\ddddot{q} = \ddddot{q}( \dddot{q}, \ddot{q}, \dot{q},q), \label{ortroNonDeg0}
\end{equation}
or what is the same, the solution depends on four quantities of initial data
\begin{equation}
q = q(\dddot{q}_0, \ddot{q}_0, \dot{q}_0,q_0,t). \label{ortroNonDeg}
\end{equation}
This solution indicates the existences of four canonical variables in the phase space. Ostrogradsky \cite{Ostrogradsky:1850fid} propose to define these variables as
\begin{align}
Q_1 \equiv q, \qquad 
Q_2 \equiv \dot{q}, \qquad 
P_1 \equiv \frac{\partial L}{\partial \dot{q}} - \frac{d}{dt} \frac{\partial L}{\partial \ddot{q}}, \qquad
P_2 \equiv \frac{\partial L}{\partial \ddot{q}}\;\;.
\end{align}
The nondegeneracy of the Lagrange function implies that $\ddot{q}$ can be solved in terms of $\ddot{q} = A(Q_1, Q_2, P_2)$ excluding momentum $P_1$, which is only needed for the third derivative.

Ostrogradsky's Hamiltonian is obtained using the Legendre transformation
\begin{equation}
H(Q_1,Q_2, P_1, P_2) = P_1 Q_2 + P_2 A(Q_1,Q_2, P_2) - L(Q_1,Q_2,A(Q_1,Q_2, P_2). \label{OrtoHamiltonian}
\end{equation}
The Hamilton equations are
\begin{align}
\dot{Q}_1 &= \frac{\partial H}{\partial P_1} = Q_2
\\
\dot{Q}_2 &= \frac{\partial H}{\partial P_2} = A + P_2 \frac{\partial A}{\partial P_2} -
\frac{\partial L}{\partial \ddot{q}} \frac{\partial A}{\partial P_2} = A
\\
\dot{P}_2 &= - \frac{\partial H}{\partial Q_2} = -P_1 -P_2 \frac{\partial A}{\partial Q_2}
+ \frac{\partial L}{\partial \dot{q}} 
+ \frac{\partial L}{\partial \ddot{q}} \frac{\partial A}{\partial Q_2}
= -P_1 + \frac{\partial L}{\partial \dot{q}}
\\
\dot{P}_1 &= - \frac{\partial H}{\partial Q_1} =  -P_2 \frac{\partial A}{\partial Q_1} 
+ \frac{\partial L}{\partial q} 
+ \frac{\partial L}{\partial \ddot{q}} \frac{\partial A}{\partial Q_1}
= \frac{\partial L}{\partial q} 
\end{align}
The first two equations reproduce the phase space transformation $\dot{q} = Q_2$ and $\ddot{q} = \dot{Q}_2$ while the others show that the evolution of momentum $P_1$ depends on the evolution of $P_2$. The equations exhibit the time evolution of the system generated by Ostrogradsky's Hamiltonian. Lagrangian is also the conserved Noether current when it contains no explicit dependence of time \cite{Woodard:2015zca}.

The Ostrogradsky's Hamiltonian \ref{OrtoHamiltonian} is linear in the canonical momentum $P_1$, which means that there is instability. The instability also is manifested because the Lagrangian depends on fewer coordinates than the defined canonical coordinates. The method mixtures the derivatives $\frac{\partial^{(n)} L}{\partial q^{(n)}}$ because the new canonical variables involve combinations of values $\dddot{q}_0, \ddot{q}_0, \dot{q}_0,q_0$ in equation \ref{ortroNonDeg}. This feature unnecessarily obscures the physical meaning of the future canonical variables in systems and entangle the quantization of the phase space structure.

The Hamiltonian of a nondegenerate higher derivative theory obtained by Ostrogradsky's is unbounded below, and above. The Ostrogradsky's instability implies that \cite{Woodard:2015zca}:
\begin{itemize}
\item The dynamical variable is provided with a special time dependence
\item Same higher derivative dynamical variable carries both positive and negative energy creation and annihilation operators. 
\item  Empty states can decay into a set of positive and negative energy excitation. One consequence of this is that vacuum can decay!!
\item In the continuum field theory, the vast entropy at infinite 3-momentum will make the decay instantaneous
\item Degrees of freedom with large 3-momentum do not decouple from low energy physics on interacting systems.
\end{itemize}
A single, global constraint on the energy functional is insufficient to mitigate the effects of the Ostrogradski instability.

\section{The second order Hamilton equations}
We propose a different construction than Ostrogradski and start by writing the total time derivative of a general second-order Lagrangian:
\begin{equation}
\frac{dL}{dt} = \sum_i \frac{\partial L}{\partial q_i} \frac{d q_i}{dt} +  \frac{\partial L}{\partial \dot{q}_i} \frac{d\dot{q}_i}{dt} +  \frac{\partial L}{\partial \ddot{q}_i} \frac{d \ddot{q}_i}{dt} +  \frac{\partial L}{\partial t}. 
\end{equation} 
Including the second order Lagrange equation \cite{Courant53physics}
\begin{equation*}
\frac{d^2}{d t^2}\Big( \frac{\partial L}{\partial \ddot{q}_i}\Big) - \frac{d}{dt}\Big( \frac{\partial L}{\partial \dot{q}_i}\Big) + \frac{\partial L}{\partial q_i} =0,
\end{equation*}
we obtain
\begin{equation}
\frac{dL}{dt} = \sum_i \Big[ \frac{d}{dt} \Big( \frac{\partial L}{\partial \dot{q}_i}\Big) - \frac{d^2}{dt^2}\Big( \frac{\partial L}{\partial \ddot{q}_i} \Big) \Big] \dot{q}_i+ \frac{\partial L}{\partial \dot{q}_i} \frac{d\dot{q}_i}{dt} +  \frac{\partial L}{\partial \ddot{q}_i} \frac{d \ddot{q}_i}{dt} +  \frac{\partial L}{\partial t}.
\end{equation}
After some derivative steps we have:
\begin{equation}
\frac{d L}{dt} = \sum_i \frac{d}{dt} \Big[ \frac{\partial L}{\partial \dot{q}_i} \dot{q}_i - \frac{d}{dt}  \Big( \frac{\partial L}{\partial \ddot{q}_i}\Big) \dot{q}_i + \frac{\partial L}{\partial \ddot{q}_i}\ddot{q}_i \Big]+\frac{\partial L}{\partial t}
\end{equation}
or
\begin{equation}
\sum_i \frac{d}{dt} \Big[ \frac{\partial L}{\partial \dot{q}_i} \dot{q}_i - \frac{d}{dt}  \Big( \frac{\partial L}{\partial \ddot{q}_i}\Big) \dot{q}_i + \frac{\partial L}{\partial \ddot{q}_i}\ddot{q}_i  - L \Big] +\frac{\partial L}{\partial t} = 0.
\end{equation}
We can define an second order energy function \\ $h(t, q_1,q_2...q_n,\dot{q}_1,\dot{q}_2...\dot{q}_n,\ddot{q}_1,\ddot{q}_2...\ddot{q}_n)$ as:
\begin{equation}
h= \frac{\partial L}{\partial \dot{q}_i} \dot{q}_i - \frac{d}{dt}  \Big( \frac{\partial L}{\partial \ddot{q}_i}\Big) \dot{q}_i + \frac{\partial L}{\partial \ddot{q}_i}\ddot{q}_i  - L
\end{equation}
where
\begin{equation}
\frac{dh}{dt}=-\frac{\partial L}{\partial t}.
\end{equation}
If Lagrangian doesn't explicit depend on time, then $h$ will remain constant in time. We can define the first and second order momenta
\begin{equation}
p_i=\frac{\partial L}{\partial \dot{q}_i} \;\;\; \text{and} \;\;\;\; s_i=\frac{\partial L}{\partial \ddot{q}_i}
\end{equation}
so the energy function is written as
\begin{equation}
h= \sum_i p_i \dot{q}_i - \dot{s}_i\dot{q}_i + s_i\ddot{q}_i  - L. \label{extEnergyFunction}
\end{equation}
The previous definitions for the canonical variables provide a clear structure of the phase space, hence, a simplified underlying symplectic geometry.

If we compute the total differential for the energy function
\begin{align}
dh = &\sum_i p_i d\dot{q}_i +  \dot{q}_i  dp_i - \dot{s}_id\dot{q}_i - \dot{q}_id\dot{s}_i  + s_id\ddot{q}_i + \ddot{q}_ids_i  
\nonumber \\
&- \Big[\frac{\partial L}{\partial q_i} dq_i + \frac{\partial L}{\partial \dot{q}_i} d\dot{q}_i + \frac{\partial L}{\partial \ddot{q}_i} d\ddot{q}_i + \frac{\partial L}{\partial t} dt\Big]
\end{align}
and substitute the momentums $p$ and $s$ definitions, we obtain
\begin{equation}
dh + \sum_i d(\dot{s}_i\dot{q}_i) = \sum_i -(\dot{p}_i-\ddot{s}_i) dq_i +  \dot{q}_i  dp_i  + \ddot{q}_i ds_i -  \frac{\partial L}{\partial t} dt,
\end{equation}
were we use the relation
\begin{equation}
\frac{\partial L}{\partial q_i}= \frac{d}{dt} \Big( \frac{\partial L}{\partial \dot{q}_i}\Big) - \frac{d^2}{dt^2}\Big( \frac{\partial L}{\partial \ddot{q}_i} \Big)=\dot{p}_i-\ddot{s}_i.
\end{equation}
We can define the function
\begin{equation}
H = h + \sum_i \dot{s}_i\dot{q}_i = \sum_i p_i \dot{q}_i + s_i\ddot{q}_i  - L. \label{ExtHamiltonDef}
\end{equation}
as a function depending only on variables $q,p,s$, whose differential is
\begin{equation}
d H = \sum_i -(\dot{p}_i-\ddot{s}_i)dq_i +   \dot{q}_i  dp_i + \ddot{q}_ids_i -  \frac{\partial L}{\partial t} dt.
\end{equation}
The differential of function $H$ can be also written as
\begin{equation}
dH = \sum_i \frac{\partial H}{\partial q_i} dq_i + \frac{\partial H}{\partial p_i} dp_i + \frac{\partial H}{\partial s_i} ds_i + \frac{\partial H}{\partial t} dt
\end{equation}
from we obtain $3n+1$ relations:
\begin{align}
&\frac{\partial H}{\partial q_i} =  -(\dot{p}_i-\ddot{s}_i)
\nonumber \\
&\frac{\partial H}{\partial p_i} = \dot{q}_i
\nonumber \\
&\frac{\partial H}{\partial s_i} = \ddot{q}_i
\nonumber \\
&\frac{\partial H}{\partial t} = -  \frac{\partial L}{\partial t} \label{ExtHamiltonEq}
\end{align}
Mathematically speaking, the set of $3n+1$ second-order second order Hamilton equations replace the set of $n+1$ four-order second order Lagrange equations.

We are in the presence of a problem where the energy function is different from Hamiltonian. This issue is in contrast to the classical mechanic where both functions are the same if there is no explicit time dependence on Hamiltonian. In that case, the energy function is the energy of the system, while Hamiltonian is the generator of the evolution of the system with time.

Since index summation $n$ stands for particle iteration on both second-order Lagrangian and Hamiltonian, we can also define the particle energy function as:
\begin{equation}
h_n = H_n - \dot{s}_{n}\dot{q}_{n}. \label{extPartEnergyFunction}
\end{equation}
In this case, particle energies are no longer constant with time. Only its summation over all particles remains invariant.

We summarize the so far obtained equations in the table \ref{extSummaryTable}. Note that the present proposal for the second order Hamiltonian has $n$ more degrees of freedom that the Lagrange approach, which resembles the Ostrogradsky's instability. This fact means that another set of equations is needed for the correct description of the system. The requested equations are obtained in the next sections.
\begin{table}[h]
\caption{Summary of the number of equations, variables and degree of freedom for the second order classical theory for Lagrange and Hamilton} \label{extSummaryTable}
\begin{center}
\begin{tabular}{@{} p{15mm}  p{5cm} p{4cm}  p{4cm} @{}}
\hline \hline
 & Equations  & Variables & Degree of freedom    \\ 
\hline
Lagrange
&
$n$ equation:
\newline
$
\frac{d^2}{dx^2}\Big( \frac{\partial L}{\partial \ddot{q}_i}\Big) - \frac{d}{dx}\Big( \frac{\partial L}{\partial \dot{q}_i}\Big) + \frac{\partial L}{\partial q_i} =0
$
& 
5-$n$ +1 initial values:
\newline
$\ddddot{q}_{i_0},\dddot{q}_{i_0}, \ddot{q}_{i_0}, \dot{q}_{i_0},q_{i_0},t$
& 5$n$+1 - 1$n$ = 4$n$+1\\ 
\hline
Hamilton
&
$3n$ equations:
\newline
$
\begin{array} {r@{} l@{}}
\dfrac{\partial H}{\partial q_i} {}&=  -(\dot{p}_i-\ddot{s}_i)
\\ 
\dfrac{\partial H}{\partial p_i} {}&= \dot{q}_i
\\ 
\dfrac{\partial H}{\partial s_i} {}&= \ddot{q}_i
\end{array}$
& 
8-$n$ +1 initial values:
\newline
$\ddot{q}_{i_0}, \dot{q}_{i_0},q_{i_0}, \dot{p}_{i_0},p_{i_0}, 
\newline
\ddot{s}_{i_0}, \dot{s}_{i_0},s_{i_0},t$
& 8$n$+1 - 3$n$ = 5$n$+1\\ 
\hline
\end{tabular} 
\end{center}
\end{table}

\section{Canonical transformations}
The Hamilton theory's basics concepts have an essential role in the construction of modern theories as quantum mechanics. One of them is the canonical transformation, which is the base to determine one of the main components in the modern quantum formalism: the operator. After obtaining the second order Hamilton equations, we will be able then to define the canonical transformations for $n$-VMVF systems depending on the second-order derivative of generalized coordinates $\ddot{q}_i$. 

Canonical transformations are said to be the standard transformations of the system going from one set of coordinates to another while the second order Hamilton equations \ref{ExtHamiltonEq} are preserved. Under the Hamiltonian formulation, the transformation of the system involves the simultaneous changes of the variables $q_i$, $p_i$ and $s_i$ into a new set $Q_i$, $P_i$ and $S_i$  with the following (invertible) transformations equations:
\begin{align}
Q_i&=Q_i(q_i,p_i,s_i) \nonumber \\
P_i&=P_i(q_i,p_i,s_i) \nonumber \\
S_i&=S_i(q_i,p_i,s_i) \label{newCordOldCord}
\end{align}
where $Q_i$, $P_i$ and $S_i$ satisfy:
\begin{align}
&\frac{\partial K}{\partial Q_i} =  -(\dot{P}_i-\ddot{S}_i)
\nonumber \\
&\frac{\partial K}{\partial P_i} = \dot{Q}_i
\nonumber \\
&\frac{\partial K}{\partial S_i} = \ddot{Q}_i \label{ExtHamiltonEqNew}
\end{align}
being $K$ the new transformed Hamiltonian. The transformation may include a factor $\lambda$ which describes a more global transformation known as "scale transformation." Here we assume $\lambda = 1$.

The function $K$ must also satisfy the least action principle:
\begin{equation}
\delta \int_{t_0}^{t_1} L(Q_i, \dot{Q}_i, \ddot{Q}_i) dt = \delta \int_{t_0}^{t_1} \sum_i P_i \dot{Q}_i + S_i \ddot{Q}_i - K(\bar{Q}_i, \bar{P}_i, \bar{S}_i,t)dt = 0
\end{equation}
where the bars symbols stand for the group of variables. Also, $H$, $q_i$, $p_i$ and $s_i$ satisfy:
\begin{equation}
\delta \int_{t_0}^{t_1} L(q_i, \dot{q}_i, \ddot{q}_i) dt = \delta \int_{t_0}^{t_1} \sum_i p_i \dot{q}_i + s_i \ddot{q}_i - H(\bar{q}_i, \bar{p}_i, \bar{s}_i,t)dt = 0.
\end{equation}
Both integrand are not equals. Instead they are connected by the relation:
\begin{equation}
 \sum_i p_i \dot{q}_i + s_i \ddot{q}_i - H(\bar{q}_i, \bar{p}_i, \bar{s}_i,t) = \sum_i P_i \dot{Q}_i + S_i \ddot{Q}_i - K(\bar{Q}_i, \bar{P}_i, \bar{S}_i,t) + \frac{dF}{dt} \label{oldNewHamilrelations}
\end{equation}
where $F$ is any function depending on the coordinates of the phase space with continuous second derivatives. The contribution of function $F$ to the variation of the action integral occurs only at the endpoints. The time derivative
\begin{equation}
\int_{t_1}^{t_2} \frac{dF}{dt} \; dt = F(2)- F(1)
\end{equation}
shows that if the function $F$ depends on the old and the new canonical variables, its variation is zero since canonical variables have zero variations at the endpoints. 

The relations \ref{newCordOldCord} connect the old and the new coordinates then, function $F$ shall depend on a combination of such type of coordinates up to the total value of $3n$. Let's suppose that the transformation function has the $F_1(q,Q,S)$ dependency. We can introduce, with no loss of generality, $2n$ more variables - $\dot{q}$ and $\dot{Q}$ -  to  $F_1$ function. Its dependency now is $F_1(q,\dot{q},Q,\dot{Q},S)$. $\dot{q}$ and $\dot{Q}$ variables are not independent on function $F$, in fact, we need $2n$ more relations for function $F_1(q,Q,S)$ keep its original $3n$ variables $(q,Q,S)$. We have $n$ relations from the straight time derivative of relations \ref{newCordOldCord}:
\begin{align}
&\dot{Q}_i=\frac{\partial Q_i}{\partial q_i}\dot{q}_i + \frac{\partial Q_i}{\partial p_i}\dot{p}_i+ \frac{\partial Q_i}{\partial s_i}\dot{s}_i
\nonumber \\
&\dot{P}_i=\frac{\partial P_i}{\partial q_i}\dot{q}_i + \frac{\partial P_i}{\partial p_i}\dot{p}_i+ \frac{\partial P_i}{\partial s_i}\dot{s}_i
\nonumber \\
&\dot{S}_i=\frac{\partial S_i}{\partial q_i}\dot{q}_i + \frac{\partial S_i}{\partial p_i}\dot{p}_i+ \frac{\partial P_i}{\partial S_i}\dot{s}_i.
\label{newCordOldCordDiff}
\end{align}
Others $n$ relations will be obtained later in the study of the identity transformation. 

Substituting $F_1(q,\dot{q},Q,\dot{Q},S)$ in equation \ref{oldNewHamilrelations} we obtain:
\begin{align}
\sum_i p_i \dot{q}_i + s_i \ddot{q}_i - H &= \sum_i P_i \dot{Q}_i + S_i \ddot{Q}_i -K
\nonumber \\
& + \frac{\partial F_1}{\partial q_i} \dot{q}_i + \frac{\partial F_1}{\partial \dot{q}_i} \ddot{q}_i + \frac{\partial F_1}{\partial Q_i} \dot{Q}_i + \frac{\partial F_1}{\partial \dot{Q}_i} \ddot{Q}_i + \frac{\partial F_1}{\partial S_i} \dot{S}_i +  \frac{\partial F_1}{\partial t} \label{F1HKrelation}
\end{align}

Since the old and new coordinates are separately independent, the equation holds if each coefficient of $\dot{q}_i$, $\ddot{q}_i$, $\dot{Q}_i$ and $\ddot{Q}_i$ vanish, from where we obtain:
\begin{align}
&p_i = \frac{\partial F_1}{\partial q_i}
,\;\;\;\; 
s_i = \frac{\partial F_1}{\partial \dot{q}_i}
,\;\;\;\;
P_i = -\frac{\partial F_1}{\partial Q_i}
,\;\;\;\;
S_i = - \frac{\partial F_1}{\partial \dot{Q}_i}
,\;\;\;\;
0 = \frac{\partial F_1}{\partial S_i}
\nonumber \\
&K=H+ \frac{\partial F_1}{\partial t}
\end{align}

Another transformation can be a different function depending on the new momentum $P$ as $F_2(q,P,S)$. We can obtain the new function $F_2$ from function $F_1$ using the D'Alembert transformation as
\begin{equation}
F_1 = F_2 - Q_i P_i
\end{equation}
We can also expand the $F_2(q,P,S)$ function with the variables $\dot{q}$ and $\dot{Q}$ to the function $F_2(q,\dot{q},P,\dot{Q},S)$, being $\dot{q}_i$ and $\dot{Q}_i$, not independent variables. Again we will need 2-$n$ more relations for the added variables, so the former function $F_2$ depends only on the $3n$ initials variables as $F_2(q,\dot{q},P,\dot{Q},S)$. Same as the previous case, we have the $n$ relations given by the straight time derivative of the transformation relations \ref{newCordOldCord} shown in equation \ref{newCordOldCordDiff}. The others $n$ relations will be obtained once we study the Identity transformation for this type of functions.

The relation between the two Hamiltonians for this type of functions, equation \ref{F1HKrelation}, can be written as:
\begin{align}
\sum_i p_i \dot{q}_i + s_i \ddot{q}_i - H &= \sum_i -Q_i \dot{P}_i + S_i \ddot{Q}_i -K
\nonumber \\
& + \frac{\partial F_2}{\partial q_i} \dot{q}_i + \frac{\partial F_2}{\partial \dot{q}_i} \ddot{q}_i + \frac{\partial F_2}{\partial P_i} \dot{P}_i + \frac{\partial F_2}{\partial \dot{Q}_i} \ddot{Q}_i + \frac{\partial F_2}{\partial S_i} \dot{S}_i +  \frac{\partial F_2}{\partial t}.
\end{align}
The coefficient of the terms $\dot{q}_i$, $\ddot{q}_i$, $\dot{Q}_i$ and $\ddot{Q}_i$ must vanish, leading to equations:
\begin{align}
&p_i = \frac{\partial F_2}{\partial q_i}
,\;\;\;\; 
s_i = \frac{\partial F_2}{\partial \dot{q}_i}
,\;\;\;\;
Q_i = \frac{\partial F_2}{\partial P_i}
,\;\;\;\;
S_i = - \frac{\partial F_2}{\partial \dot{Q}_i}
,\;\;\;\;
0 = \frac{\partial F_2}{\partial S_i}
\nonumber \\
&K=H+ \frac{\partial F_2}{\partial t} \label{TransfEqF2}
\end{align}

We proceed now to define the identity transformation. Let us consider this canonical transformation as a $F_2$ function type. In that case, the most straightforward Identity transformation has the form 
\begin{equation}
F_2 = \sum_i q_i P_i + \dot{q}_i S_i -  s_i \dot{Q}_i \label{Ident1}.
\end{equation}
From were the vanishing coefficients of equations \ref{TransfEqF2} result in
\begin{align}
&p_i = \frac{\partial F_2}{\partial q_i} = P_i
,\quad
s_i = \frac{\partial F_2}{\partial \dot{q}_i} = S_i
,\quad
Q_i = \frac{\partial F_2}{\partial P_i} = q_i,
\nonumber \\
&S_i = - \frac{\partial F_2}{\partial \dot{Q}_i} = s_i
,\quad 
K=H
,\quad 
0 = \frac{\partial F_2}{\partial S_i} = \dot{q}_i \label{Ident3}
\end{align}
The first five equations of equations \ref{Ident3} shows that the old and the new coordinates are the same, probing function \ref{Ident1} being a suitable candidate for identity transformation. In the obtaining of the transformation equation for functions  $F_1$ and $F_2$, we added variables  $\dot{q}_i$ and $\dot{Q}_i$  as a dependent set of variables. We stated that still $n$  more relations are needed between these coordinates, so the transformation is successfully described. Last set equation of \ref{Ident3} are those wanted relations. However, the obtained set of equations, $ \dot{q}_i=0$ are not acceptable solutions to our problem.

We instead, propose the identity transformation as:
\begin{equation}
F_2 = \sum_i q_i P_i + \mathcal{F}_{2_i}(\bar{\dot{q}}) S_i -  s_i \dot{Q}_i \label{Identity}.
\end{equation}
where $\mathcal{F}_{2_i}(\bar{\dot{q}})$ is the $i$-component of a function depending of all $\{\dot{q}_i\}$ that satisfied:
\begin{equation}
\frac{\partial \mathcal{F}_{2_i}(\bar{\dot{q}})}{\partial \dot{q}_j} = \delta_{ij},\;\;\;\;\;  \mathcal{F}_{2_i}(\bar{\dot{q}})\neq \dot{q}_i  + C_i\label{correlFunctDef}
\end{equation}
being $C_i$ constants. For this transformation, we obtain the relations:
\begin{align}
&p_i = \frac{\partial F_2}{\partial q_i} = P_i
,\quad 
s_i = \frac{\partial F_2}{\partial \dot{q}_i} = \frac{\partial  \mathcal{F}_{2_i}(\bar{\dot{q}})}{\partial \dot{q}_i} S_i= S_i,
\nonumber \\
&Q_i = \frac{\partial F_2}{\partial P_i} = q_i
,\quad  
S_i = - \frac{\partial F_2}{\partial \dot{Q}_i} = s_i \label{Identity2}
,\quad 
K=H
\\
&0 = \frac{\partial F_2}{\partial S_i} = \mathcal{F}_{2_i}(\bar{\dot{q}}) \label{correlFunctCond}
\end{align}
The equations  \ref{correlFunctCond}
\begin{equation*}
 \mathcal{F}_{2_i}(\bar{\dot{q}}) = 0
\end{equation*}
are the $n$ remaining relations needed for variables $q, P, S$,  being the only independent degrees of freedom in function $F_2(q,\dot{q}, P, \dot{Q}, S)$.  Once the forms of the correlation functions $F_2(q,\dot{q},P, \dot{Q},S)$ are defined, they set $n$ relations between the generalized velocities. These new constraints reduce the number of canonical variables equals the number of the variables on the Lagrange approach, removing the Ostrogradsky's instability.

If the particle system has only one particle, the only possible solution for $\mathcal{F}_{2_i}$, according to the definition $\frac{\partial \mathcal{F}_{2_i}(\bar{\dot{q}})}{\partial \dot{q}_j} = \delta_{ij}$, is precisely the restriction we imposed:   $\mathcal{F}_{2} = \dot{q} + C$. Even, we restricted this solution, is worth analyzing the implications of it. Our proposal for the definitions of the canonical variables
We already state that $ \dot{q} = - C$ is not an allowed solution to our problem. That means that we cannot define an identity transformation for a one particle system depending on $\ddot{q}$. This result is consistent with one of the conclusion arrived on the works that precede and motivate this proposal \cite{Israel:1811.12175, Israel:1903.04916}. Indeed, the $\ddot{q}$'s dependency appears in the problem described on the referred works when mass is assumed as a variable quantity without any restriction. It is well known that this assumption leads to the noncompliance of the second Newton law for one isolated particle, because of the violation of the relativity principle under a Galilean transformation \cite{Plastino1992}. We proposed that, for an isolated particle system, such violation can be suppressed by the action of the mass's variation of the others particle. Then, an isolated particle system whose mass of the particle varies must include at least two particles or the particle. The above result reinforce these predictions.

\section{Infinitesimal canonical transformations}
We study now the infinitesimal canonical transformations were new variables differ from the old ones just by infinitesimals. In that case, the transformation equations \ref{newCordOldCord} have the form:
\begin{align}
Q_i=q_i + \delta q_i 
\nonumber \\
P_i=p_i + \delta p_i 
\nonumber \\
S_i=q_i + \delta s_i, 
\end{align}
where $\delta q_i$, $\delta p_i$ and $\delta s_i$ are the real displacements of each variable, respectively. The infinitesimal canonical transformation can be written as the sum of the identity transformation plus an infinitesimal function. In the case of transformations describe with $F_2$ type functions, they have the form
\begin{equation}
F_2 = \sum_i q_i P_i + \mathcal{F}_{2_i}(\bar{\dot{q}}) S_i -  s_i \dot{Q}_i + \epsilon \mathcal{G}(q_i, p_i, s_i, t),
\end{equation}
being $\epsilon$ an infinitesimal parameter for describing the magnitude of the transformation and $\mathcal{G}(q_i,\dot{q}_i, p_i,\dot{Q}_i, s_i, t)$ is a differentiable function with $3n+1$ arguments known as the generator of such transformation. After applying equations \ref{TransfEqF2},  we obtain the transformation relations:
\begin{align}
&p_i = \frac{\partial F_2}{\partial q_i} = P_i + \epsilon \frac{\partial \mathcal{G}}{\partial q_i}
\;\;\;\; \text{or} \;\;\;\; \delta p_i = - \epsilon \frac{\partial \mathcal{G}}{\partial q_i}
\nonumber \\
&s_i = \frac{\partial F_2}{\partial \dot{q}_i} = S_i + \epsilon \frac{\partial \mathcal{G}}{\partial \dot{q}_i}
\;\;\;\; \text{or} \;\;\;\; \delta s_i = - \epsilon \frac{\partial \mathcal{G}}{\partial \dot{q}_i}
\nonumber \\
&Q_i = \frac{\partial F_2}{\partial P_i} = q_i + \epsilon \frac{\partial \mathcal{G}}{\partial P_i}
\;\;\;\; \text{or} \;\;\;\; \delta q_i = \epsilon \frac{\partial \mathcal{G}}{\partial P_i}
\nonumber \\
&S_i = - \frac{\partial F_2}{\partial \dot{Q}_i} = s_i + \epsilon \frac{\partial \mathcal{G}}{\partial \dot{Q}_i} 
\;\;\;\; \text{or} \;\;\;\; \delta s_i = \epsilon \frac{\partial \mathcal{G}}{\partial \dot{Q}_i}
\nonumber \\
&0 = \frac{\partial F_2}{\partial S_i} = \mathcal{F}_{2_i}(\bar{\dot{q}})  + \epsilon \frac{\partial \mathcal{G}}{\partial S_i}.
\end{align}

The infinitesimal canonical transformation generated by the generalized new momentum $P_i$
\begin{equation}
\mathcal{G} = P_i,
\end{equation}
result in the coordinates variations
\begin{align}
\delta q_j = \epsilon \delta_{ij}, 
\nonumber \\
\delta p_j = 0, 
\nonumber \\
\delta s_j = 0, 
\nonumber \\
\mathcal{F}_{2_j}(\bar{\dot{q}})=0.
\end{align}
The set of equations shows that the generator $\mathcal{G} = P_i = p_i +\delta p_i = p_i$ transforms the system displacing only of coordinate $q_i$ if parameter $\epsilon$ is the displacement value. This fact settles the generalized momentum as the generator of the displacement of its own coordinate, coincident with the first order theory of Hamilton.

The transformation of the system with the new momentum $S_i$ as the generator of the transformation
\begin{equation}
\mathcal{G} = S_i,
\end{equation}
is described by the coordinates changes:
\begin{align}
\delta q_j = 0, 
\nonumber \\
\delta p_j = 0, 
\nonumber \\
\delta s_j = 0, 
\nonumber \\
\mathcal{F}_{2_j}(\bar{\dot{q}})=-\epsilon \delta_{ij}. \label{sGenEquations}
\end{align}
According to this results, the generator $s_i$ transforms the system keeping unaltered the variables $q_i,p_1$ and $s_i$, an modifying the value of the right member of the equation $\mathcal{F}_{2_i}(\bar{\dot{q}})=0$ to the infinitesimal displacement value. 

We define the new variable $f_j(\bar{\dot{q}})$ as the value of the correlation $j$-function at any time. The form of the correlation functions remain unchanged across the evolution of the system, but their value will vary as $\mathcal{F}_{2_i}(\bar{\dot{q}}) = f_j(\bar{\dot{q}})$. 

The obtained equations \ref{correlFunctCond}, show that all resulting values $f_j(\bar{\dot{q}})$ are zero in the identity transformation, $f_{0_j}(\bar{\dot{q}})=0$. Then, the last equation of \ref{sGenEquations} can be written as
\begin{equation}
\mathcal{F}_{2_j}(\bar{\dot{q}})= f_j(\bar{\dot{q}})-0 = f_j(\bar{\dot{q}}) - f_{0_j}(\bar{\dot{q}})= \delta f_j(\bar{\dot{q}})= -\epsilon \delta_{ij}.\label{corrFunctionValue}
\end{equation}

The second order momentum $s$ can be interpreted, then, as the generator of a negative displacement of the value of correlation functions, $\mathcal{F}_{2_i}(\bar{\dot{q}})$. Being the correlation function a constraint involving all particle of the system, we can conclude that the new momentum $s$ is the generator of collective action of the system. This behavior is in agreement with our initial supposition where the violation of the Newton second law, introduced by the terms proportional to $\ddot{q}$, will be suppressed by the coordinate action of all the particles of the system.

Another important canonical transformation is
\begin{equation}
\mathcal{G} = H + \sum_i \dot{s}_i\dot{Q}_i - \dot{s}_i\dot{q}_i-\ddot{s}_i q_i. \label{timeGenerator}
\end{equation}
The infinitesimal changes in the variables of the system are
\begin{align}
&\delta p_i = - \epsilon \frac{\partial \mathcal{G}}{\partial q_i}= -\epsilon \Big(\frac{\partial H}{\partial q_i} -\ddot{s}_i \Big ) = \epsilon \dot{p}_i
\nonumber \\
&\delta s_i = - \epsilon \frac{\partial \mathcal{G}}{\partial \dot{q}_i} = \epsilon \dot{s}_i
\nonumber \\
&\delta q_i = \epsilon \frac{\partial \mathcal{G}}{\partial P_i} \sim \epsilon \frac{\partial \mathcal{G}}{\partial p_i} = \epsilon \Big(\frac{\partial H}{\partial p_i} \Big) = \epsilon \dot{q}_i
\nonumber \\
&\delta s_i = \epsilon \frac{\partial \mathcal{G}}{\partial \dot{Q}_i} = \epsilon \dot{s}_i
\nonumber \\
&0 = \frac{\partial F_2}{\partial S_i} = \mathcal{F}_{2_i}(\bar{\dot{q}})  + \epsilon \frac{\partial \mathcal{G}}{\partial S_i}  \sim \mathcal{F}_{2_i}(\bar{\dot{q}})  + \epsilon \frac{\partial \mathcal{G}}{\partial s_i}  = \mathcal{F}_{2_i}(\bar{\dot{q}})  + \epsilon \frac{\partial H}{\partial s_i} 
\nonumber \\
& \mathcal{F}_{2_i}(\bar{\dot{q}}) \equiv \delta f_j(\bar{\dot{q}}) =- \epsilon \ddot{q}_i. \label{extCanonEqTime}
\end{align}
On the other side, the time derivative of the function $\mathcal{F}_{2_i}(\bar{\dot{q}}) $ is
\begin{equation}
\frac{d f_i(\bar{\dot{q}})}{dt} 
= \frac{d \mathcal{F}_{2_i}(\bar{\dot{q}})}{dt} 
=  \sum_j \frac{\partial \mathcal{F}_{2_i}(\bar{\dot{q}})}{\partial q_j}\ddot{q}_j 
= \sum_j \delta_{ij}\ddot{q}_j
= \ddot{q}_i \label{fTimeDerivative}
\end{equation}
where we use the definition of function $\mathcal{F}_{2_i}(\bar{\dot{q}})$ \ref{correlFunctDef}.
The last relation of equations \ref{extCanonEqTime} can be rewritten then as:
\begin{equation}
\delta f_j(\bar{\dot{q}}) = -  \dot{f}_j(\bar{\dot{q}}) \epsilon
\end{equation}

If parameter $\epsilon$ is the infinitesimal time interval $dt$, then the generator function of equation \ref{timeGenerator}, evolves all variables of the system $q_i,p_i,s_i$ with time and also change the value of correlation functions, $ f_j(\bar{\dot{q}})$, in the negative direction.

The negative time evolution for quantities $ f_i(\bar{\dot{q}})$ is consistent with previous results where momentum $s_i$ generate a negative displacement for the value of the correlation $i$-function. According to that, in a time interval $dt$, $s$, as part of the previous time generator,  evolve with time from value $s_{i_0}$ to $s_{i_0} + ds_i$. These $s_i$ values generate the negative displacement of the value of the correlation functions $- \delta f_{2_i}(\bar{\dot{q}})$ and $- (\delta f_{i}(\bar{\dot{q}}) + \frac{\partial (ds_i)}{\partial S_i} )$ respectively, with an effective displacement of $ d(f_{i}(\bar{\dot{q}})) \equiv -\frac{\partial (ds_i)}{\partial S_i} $ or $ d(f_{2_i}(\bar{\dot{q}})) \equiv -\frac{\partial (\dot{s}_i dt)}{\partial S_i} $. Then, being $dt$ positive, the positive evolution of momentum $s_i$ evolve the quantity $f_{i}(\bar{\dot{q}})$ negatively.

The generator \ref{timeGenerator}, expressed in the old set of coordinates has the form:
\begin{align}
\mathcal{G} &= H + \sum_i  \dot{s}_i\dot{Q}_i - \dot{s}_i\dot{q}_i - \ddot{s}_i q_i = H + \sum_i  \dot{s}_i(\dot{Q}_i - \dot{q}_i) - \ddot{s}_i q_i 
\nonumber \\
&= H+\sum_i \dot{s}_i\delta \dot{q}_i - \ddot{s}_i q_i. \label{timeGenerator1}
\end{align}
We can approach $\delta \dot{q}_i \sim \ddot{q}_i t$. In this case, the system time generator \ref{timeGenerator1} is written as

\begin{equation}
\mathcal{G} \sim H + \sum_i  \dot{s}_i \ddot{q}_i t - \ddot{s}_i q_i. 
\end{equation}
or using equation \ref{fTimeDerivative}
\begin{equation}
\mathcal{G} \sim H + \sum_i  \dot{s}_i \dot{f}_i(\bar{\dot{q}}) t - \ddot{s}_i q_i. \label{timeGeneratorApx1}
\end{equation}

\section{Final second order Hamilton equations}
With the introduction of the correlation functions as new constraints for the time derivative of the canonical variable $q_i$ and the identification of its poles as a variable that evolute with the system, we replace our former second order Hamilton equations eq. \ref{ExtHamiltonEq} by
\begin{align}
&\frac{\partial H}{\partial q_i} =  -(\dot{p}_i-\ddot{s}_i) &
&\frac{\partial H}{\partial p_i} = \dot{q}_i
\nonumber \\
&\frac{\partial H}{\partial s_i} = \dot{f}_i  &
&f_i = \mathcal{F}_{2_i}(\bar{\dot{q}})
\nonumber \\
&\frac{\partial H}{\partial t} = -  \frac{\partial L}{\partial t}.
\end{align}

\section{Conclusions}
We considered the construction of the second order Hamiltonian from a second order Lagrangian. The definition of the canonical variables is different than Ostrogradsky's, hence the difference with the Hamilton equations. However, it provides a clean structure of the phase space and a simplified underlying symplectic geometry. The new definition still reproduces Ostrogradsky instability. It is the Identity canonical transformation that reveals the existence of $n$ constraints depending only on the generalized velocities. The form of the correlation functions is fixed, and they should be chosen according to the studied system. The set of constraints removes the Ostrogradsky instability as the number of variables in the phase space matches the number of variables of the configuration space.

The canonical transformations of second order Hamiltonians show that the generalized linear momentum remains as the generator of the displacement of the generalized coordinate, while the new momentum $s$ is the generator of a negative displacement of the pole of the correlation functions.

\section{acknowledgments}
I would like to express my deep gratitude to my mentor and personal friend Professor Dr. Fernando Guzm\'an Mart\'inez, for their guidance, encouragement, and critiques. I like to acknowledge professor Dr. A. Deppman for his teaching, advice and comments on this work. I also recognize the support from Dr. Yoelvis Orozco, Dr. Juan A. Garc\'ia, Dr. Yansel Guerrero and Dr. Rodrigo Gester.
\bibliography{References}

\end{document}